\documentclass[superscriptaddress,aps,prl,twocolumn,showpacs,preprintnumbers,amsmath,amssymb,floatfix,english]{revtex4}

\usepackage{amsmath}
\usepackage{amssymb}
\usepackage{graphicx}

\usepackage[colorlinks,bookmarks=false,citecolor=blue,linkcolor=red,urlcolor=blue]{hyperref}

\makeatletter

\@ifundefined{textcolor}{}
{%
 \definecolor{BLACK}{gray}{0}
 \definecolor{WHITE}{gray}{1}
 \definecolor{RED}{rgb}{1,0,0}
 \definecolor{GREEN}{rgb}{0,1,0}
 \definecolor{BLUE}{rgb}{0,0,1}
 \definecolor{CYAN}{cmyk}{1,0,0,0}
 \definecolor{MAGENTA}{cmyk}{0,1,0,0}
 \definecolor{YELLOW}{cmyk}{0,0,1,0}
 }

\@ifundefined{definecolor}
 {\usepackage{color}}{}

\usepackage[normalem]{ulem}

\usepackage{babel}

\begin{document}

\title{Antiferromagnetic spin-$S$ chains with exactly dimerized ground states}

\author{Fr\'ed\'eric Michaud}
\affiliation{Institute of Theoretical Physics, Ecole Polytechnique F\'ed\'erale de Lausanne (EPFL), CH-1015 Lausanne, Switzerland}
\author{Fran\c{c}ois Vernay}
\affiliation{Laboratoire PROMES (UPR-8521) \& UPVD, Perpignan, F-66860 Perpignan, France}
\author{Salvatore R. Manmana}
\affiliation{JILA (University of Colorado and NIST), and Department of Physics, CU Boulder, CO 80309-0440, USA}
\author{Fr\'ed\'eric Mila}
\affiliation{Institute of Theoretical Physics, Ecole Polytechnique F\'ed\'erale de Lausanne (EPFL), CH-1015 Lausanne, Switzerland}

\date{\today}
\begin{abstract}
We show that spin $S$ Heisenberg spin chains with an additional three-body interaction of the form
$(\mathbf{S_{i-1}} \cdot \mathbf{S_{i})}(\mathbf{S_{i}} \cdot \mathbf{S_{i+1}})+h.c.$ possess fully dimerized ground
states if the ratio of the three-body interaction to the bilinear one is equal to $1/(4S(S+1)-2)$.
This result generalizes the Majumdar-Ghosh point of the $J_1-J_2$ chain, to which
the present model reduces for $S=1/2$. For $S=1$, we use the density matrix
renormalization group method (DMRG) to show that the transition between the Haldane
and the dimerized phases is continuous with central charge $c=3/2$. Finally, we show
that such a three-body interaction appears naturally in a strong-coupling expansion of
the Hubbard model, and we discuss the consequences for the dimerization of actual
antiferromagnetic chains.
\end{abstract}

\maketitle

\textit{Introduction --}
Over the years, exact results have proved to be extremely useful in quantum and statistical physics \cite{sutherland_book,baxter}.
In quantum magnetism, the Bethe ansatz solution of the spin-1/2 Heisenberg chain \cite{Bethe} has led to the first proof that the spectrum is gapless \cite{des_cloizeaux}, and its extensions, e.g., to the $S=1$ chain with bilinear and biquadratic interactions (BLBQ) with equal \cite{uimin,lai,sutherland} or opposite \cite{takhtajan,babujian} amplitudes has helped a lot to clarify the physics of that model.
In quantum frustrated magnetism \cite{frustratedQMbook}, cases where an exact expression for the ground state wave function can be obtained have also played a very important role.
For instance, for the spin-1 Heisenberg chain, the exact ground state of the AKLT point \cite{AKLT} has been a milestone in the confirmation of Haldane's prediction that the spectrum of integer-$S$ spin chains is gapped \cite{haldane1983}.
For spin-1/2 magnets, the first example of a gapped spectrum goes back to the Majumdar-Ghosh \cite{MG1} (MG) point $J_2/J_1=1/2$ of the $J_1-J_2$ model defined by the Hamiltonian
\begin{equation}
\mathcal{H}_{J_1-J_2} =  \sum_i ( J_1 \, \mathbf{S}_i \cdot \mathbf{S}_{i+1} + J_2 \, \mathbf{S}_i \cdot \mathbf{S}_{i+2} ).
\label{eq:mgham}
\end{equation}
At that point, the two fully dimerized states obtained as products of singlets on consecutive dimers and defined by
\begin{equation}
|\psi_{even,odd}\rangle = \prod_{i \ even,odd} |S(i,i+1)\rangle,
\label{eq:mgstate}
\end{equation}
where $|S(i,i+1)\rangle$ denotes the singlet formed by the spins at sites $i$ and $i+1$, have been shown by Majumdar and Ghosh to be exact ground states.
Building on this result, it has been shown that the spectrum is gapped, and that this point is representative of an extended phase that covers the parameter range $0.2411 < J_2/J_1 < +\infty$ \cite{okamoto92,PhysRevB.54.R9612,PhysRevB.54.9862}.
This seminal result has been at the origin of a long series of experimental investigations of frustrated spin-1/2 chains which started about 20 years ago with CuGeO$_3$ and which remains a very active field of research \cite{Hase_PRL1993}.

Attempts at generalizing the MG point to come up with a realistic model with fully dimerized states as exact ground states for larger spins have failed so far.
The simplest idea is to consider the model of Eq.~\eqref{eq:mgham} for spins $S\geq1$ \cite{Kolezhuk_PRL1996}.
It is easy to convince oneself that the dimerized states of Eq.~\eqref{eq:mgstate} remain exact eigenstates for any spin when $J_2/J_1=1/2$, but for $S\geq1$, they are no longer the ground state.
The problem can be traced back to the properties of a single triangle, into which the Hamiltonian of Eq.~\eqref{eq:mgham} can be decomposed for $J_2/J_1=1/2$:
For $S=1/2$, the product of a singlet built out of two spins times any state of the third spin is a ground state.
For $S\geq1$, the same state has a total spin $S$, and it is not the ground state, which has total spin $0$ or $1/2$ for integer and half-integer $S$ respectively.

Following Klein \cite{Klein}, an interesting alternative consists in building Hamiltonians as sums of local projectors on three spins to ensure that the product of a singlet with a single spin state is a local ground state. The simplest Hamiltonian of that kind takes the form\cite{affleck_1989}
\begin{equation}
\mathcal{H}_{\rm Klein} = - \sum_i P_{S_{tot}=S}^{i,i+1,i+2}
\label{eq:klein}
\end{equation}
where $P_{S_{tot}=S}^{i,i+1,i+2}$ is the projector on the subspace of total spin $S$ \footnote{One can also choose any positive linear combination of projectors into states with a total spin different from S $\mathcal{H}_{\rm Klein} =  \sum_{j \neq S}\sum_i C_j P_{S_{tot} = j}^{i,i+1,i+2}$, where $C_j \geq 0$.}.
This projector can be written as
\begin{equation}
P_{S_{tot}=S}^{i,i+1,i+2}=\prod_{\sigma\neq S}\frac{(\mathbf{S}_i +\mathbf{S}_{i+1} + \mathbf{S}_{i+2})^2-\sigma(\sigma+1)}{S(S+1)-\sigma(\sigma+1)} \, ,
\label{eq:projector}
\end{equation}
where the product runs from $0$ or $1/2$ for integer or half-integer spins to $3S$.
For $S=1/2$, this Hamiltonian reduces to the MG point of the $J_1-J_2$ chain, but for $S\geq1$, it is a polynomial in scalar products of pairs of spins of degree $3S$ or $3S-1/2$ for integer or half-integer spins, hence a very complicated Hamiltonian that seems difficult to realize in actual systems. The same remark applies to a spin-3/2
model recently investigated by Rachel\cite{rachel}, whose ground states are partially dimerized valence bond solid
states, or to the generalizations proposed by Rachel and Greiter\cite{rachel_greiter} that lead to exactly trimerized
resp. tetramerized ground states for $S=1$ resp. $S=3/2$ models.

In this Letter, we propose another generalization to arbitrary $S$ of the spin-1/2 $J_1-J_2$ model defined by the Hamiltonian
\begin{equation}
\mathcal{H} =  \sum_i \left( J_1 \, \mathbf{S}_i \cdot \mathbf{S}_{i+1} + J_3  \left[\left({\bf S}_{i-1}\cdot{\bf S}_{i}\right)\left({\bf S}_{i}\cdot{\bf S}_{i+1}\right)+h.c.\right] \right)
\label{eq:J1J3ham}
\end{equation}
with $J_1 > 0$. The number of sites $N$ is assumed to be even, and we concentrate on periodic boundary conditions \footnote{The results can be extended to open boundary conditions, but then there is only one exact dimerized ground state.}.
As we shall see, this Hamiltonian possesses for any
value of $S$ the equivalent of a MG point when $J_3/J_1=1/(4S(S+1)-2)$, at which the states of Eq.~\eqref{eq:mgstate} are exact ground states, and it is realistic in the sense that it appears to next-to-leading order in the $1/U$ expansion of the two-band Hubbard model that leads to the $S=1$ Heisenberg model.

For $S=1/2$, it is easy to check that the Hamiltonian of Eq.~\eqref{eq:J1J3ham} reduces to that of Eq.~\eqref{eq:mgham} with 
$J_2=J_3/2$.
For $S\geq1$, the three-spin interaction does not reduce to a next-nearest neighbor two-spin interaction, and the proof that the states of Eq.~\eqref{eq:mgstate} are exact eigenstates is not a trivial extension of the MG proof.

As in the $S=1/2$ case, let us first determine under which condition the states of Eq.~\eqref{eq:mgstate} might
be exact eigenstates of Eq.~\eqref{eq:J1J3ham}.
To be specific, let us consider $|\psi_{odd}\rangle$.
For $i$ odd, ${\bf S}_{i}\cdot{\bf S}_{i+1}|\psi_{odd}\rangle=-S(S+1)|\psi_{odd}\rangle$.
By contrast, for $i$ even, the singlets on bonds $(i-1,i)$ and $(i+1,i+2)$ are affected by ${\bf S}_{i}\cdot{\bf S}_{i+1}$.
However, the resulting wave function does not contain states with arbitrary spin for the pairs $(i-1,i)$ and $(i+1,i+2)$, but only triplets.
Indeed, for two spins ${\bf S}_{1}$ and ${\bf S}_{2}$, $S_1^\alpha|S(1,2)\rangle$ is a triplet for all spin components $\alpha=x,y,z$.
This is clear for the $z$ component since the SU(2) commutation relations imply that
\begin{eqnarray}
(S_1^-+S_{2}^-)^2 S_1^z |S(1,2)\rangle&=&0,\nonumber \\
(S_1^++S_{2}^+)(S_1^-+S_{2}^-) S_1^z |S(1,2)\rangle&=&2S_1^z |S(1,2)\rangle,
\nonumber
\end{eqnarray}
and by rotational symmetry, this has to be true of the other components as well.
So, for $i$ even, one can write
\begin{equation*}
\begin{array}{l}
 {\bf S}_{i}\cdot{\bf S}_{i+1}|\psi_{odd}\rangle=\\[2mm]
\displaystyle \sum_{\sigma,\sigma'} C_{\sigma,\sigma'} |T_\sigma(i-1,i)\rangle |T_{\sigma'}(i+1,i+2)\rangle {\prod_{j \ odd}} ' |S(j,j+1)\rangle
\end{array}
\end{equation*}
where the product over $j$ is limited to $j\neq i-1,i+1$,
and where the indices $\sigma,\sigma'=0,\pm1$ keep track of the three possible triplets of a pair of spins.
Since the total wave-function is a singlet, all coefficients must be equal to zero except $C_{1,-1}$, $C_{-1,1}$ and $C_{0,0}$, which must be related by $C_{1,-1}=C_{-1,1}=-C_{0,0}$. Their common absolute value can be derived with the help of Clebsch-Gordan coefficients, but this is unimportant for our present purpose.
The only relevant fact is that, since only triplets are involved, acting with ${\bf S}_{i-1}\cdot{\bf S}_{i}$ or ${\bf S}_{i+1}\cdot{\bf S}_{i+2}$ on ${\bf S}_{i}\cdot{\bf S}_{i+1}|\psi_{odd}\rangle$ will just multiply it by $1-S(S+1)$.
This leads to:
\begin{eqnarray}
&&\mathcal{H}|\psi_{odd}\rangle=-\frac{J_1 N}{2}S(S+1)|\psi_{odd}\rangle\nonumber\\
&&+(J_1 -(4S(S+1)-2)J_3)\sum_{i \ even} {\bf S}_{i}\cdot{\bf S}_{i+1}|\psi_{odd}\rangle
\end{eqnarray}
If $J_3/J_1 =1/(4S(S+1)-2)$, the second term drops, and $|\psi_{odd}\rangle$ is an eigenstate of $\mathcal{H}$ with energy per site $-J_1 S(S+1)/2$.
Since the Hamiltonian is translationally invariant, this is also true for $|\psi_{even}\rangle$.

\begin{figure}[t]
\includegraphics[width=0.5\textwidth]{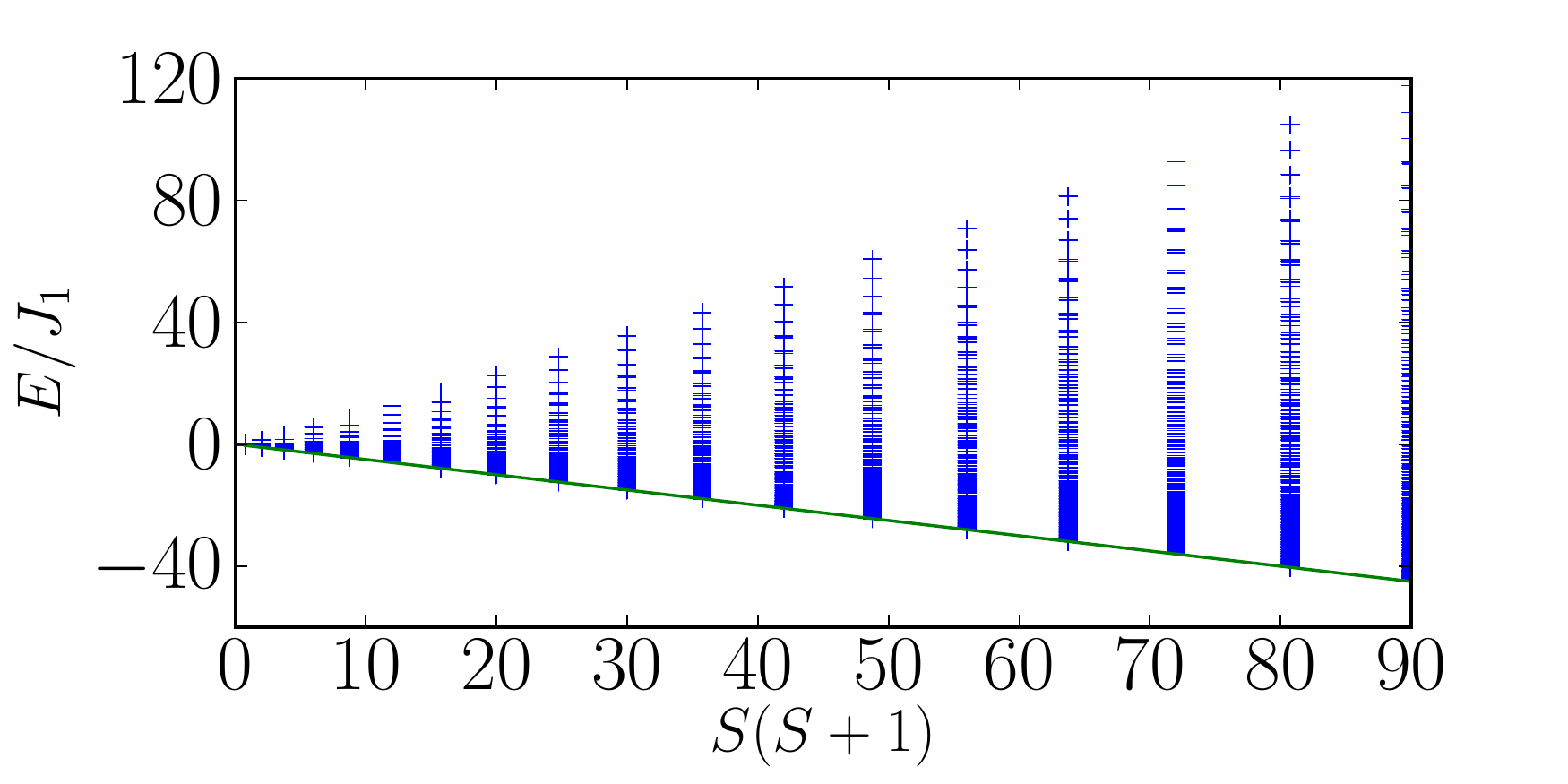}
\caption{(Color online) Spectrum of the Hamiltonian $\mathcal{H}_i$ [Eq.~\eqref{eq:3sites}] on three adjacent sites of the chain as a function of $S(S+1)$. The green line indicates the energy of the dimerized eigenstate on this three site system, $E = -S(S+1)/2$.}
\label{fig:spectre}
\end{figure}

To prove that these states are the ground states, let us decompose the Hamiltonian as $\mathcal{H}  =  J_1 \sum_{i}\mathcal{H}_i$ with
\begin{equation}
\begin{split}
\mathcal{H}_i & =\frac{1}{2}({\bf S}_{i-1}\cdot{\bf S}_{i}+{\bf S}_{i}\cdot{\bf S}_{i+1})\\
   +  & \frac{1}{4S(S+1)-2}\left[\left({\bf S}_{i-1}\cdot{\bf S}_{i}\right)\left({\bf S}_{i}\cdot{\bf S}_{i+1}\right)+h.c.\right].
\end{split}
\label{eq:3sites}
\end{equation}
The spectrum of this three-spin Hamiltonian can be worked out analytically for $S=1$ and numerically
for larger spin, with the result that the ground state energy $E_{GS}(\mathcal{H}_i)$ is equal to $-S(S+1)/2$
(see Fig.~\ref{fig:spectre}). By the variational principle, $\langle \mathcal{H} \rangle \geq J_1  \sum_i E_{GS}(\mathcal{H}_i)=-NJ_1 S(S+1)/2$, a lower bound saturated by $|\psi_{odd}\rangle$ and $|\psi_{even}\rangle$.
This completes the proof that they are ground states of the Hamiltonian of Eq.~\eqref{eq:J1J3ham} when $J_3/J_1 =1/(4S(S+1)-2)$.

Finally, it is plausible that these are the only ground states since the only ground states
of $\mathcal{H}_i$ are the wavefunctions with a singlet $\vert S(i-1,i)\rangle$ or $\vert S(i,i+1)\rangle$, and
the only common eigenstates are given by $|\psi_{odd}\rangle$ and $|\psi_{even}\rangle$. However, a
mathematically rigorous proof that these are the only ground states for infinite systems would require an
analysis similar to that of Ref.[\onlinecite{AKLT}] for the MG point of the spin-1/2 $J_1-J_2$ model.

\begin{figure}[t]
\includegraphics[width=0.5\textwidth]{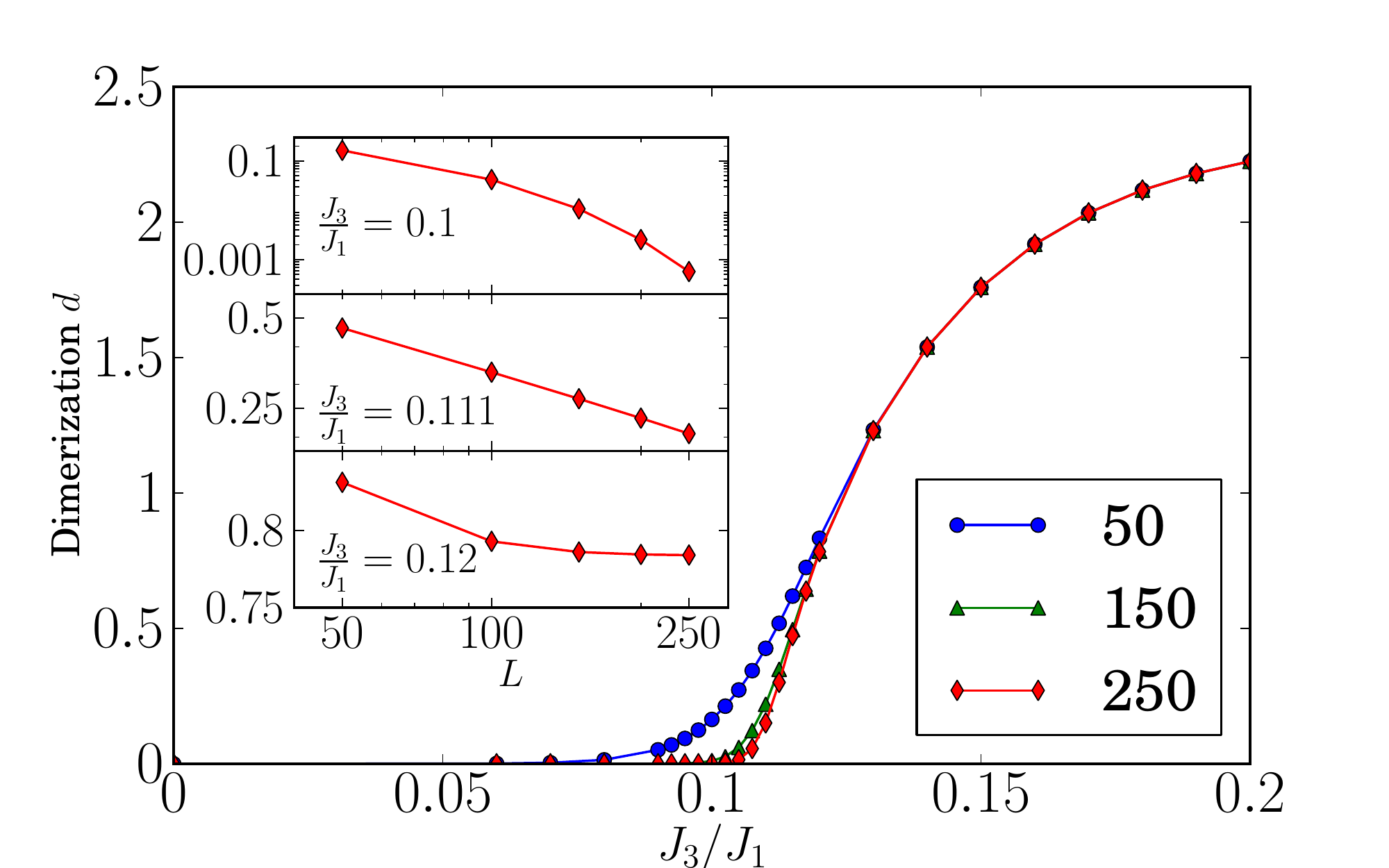}
\caption{(Color online) Dimerization as a function of $J_3/J_1$ for different system sizes up to $L=250$ sites in the vicinity of the phase transition $J_3/J_1 \approx 0.11$. The insets show the size dependence at $J_3/J_1 = 0.1, \, 0.111$ and $0.12$, respectively.}
\label{fig:dimerization}
\end{figure}

\textit{Vicinity of the MG point for $S=1$ -- } We now concentrate on the $S=1$ model.
At the Heisenberg point $J_3 = 0$, the system is in the Haldane phase, which is gapped but not dimerized.
Therefore, a phase transition has to appear between the MG point $J_3/J_1  = 1/6$ and the Heisenberg point.
Let us investigate the nature of this transition numerically using the DMRG \cite{White:1992p2171, schollwoeck2005}.

The natural order parameter of this transition is the dimerization operator defined by $d=|\langle{\bf S}_{i}\cdot{\bf S}_{i+1}-{\bf S}_{i}\cdot{\bf S}_{i-1}\rangle|$ where $(i,i+1)$ is the central bond. Results for sizes up to 250 sites\footnote{The Dimerization was computed with OBC. Close to the transition, we keep up to 1400 states and performed 9 sweeps. This keeps the discarded weight in the last sweep below $10^{-11}$} are shown in Fig.~\ref{fig:dimerization}. At the MG point, $d$ is exactly equal to 2 for all sizes.
The dimerization develops around $J_3/J_1 = 0.11$ in a way typical of a continuous transition.
Assuming this to be the case, we have performed a finite-size scaling in the vicinity of the critical point, and we have identified the point where the dimerization decays to zero algebraically.
This occurs at $J_3/J_1 = 0.111(1)$ (middle panel of the inset of Fig. ~\ref{fig:dimerization}).

\begin{figure}
\includegraphics[width=0.5\textwidth]{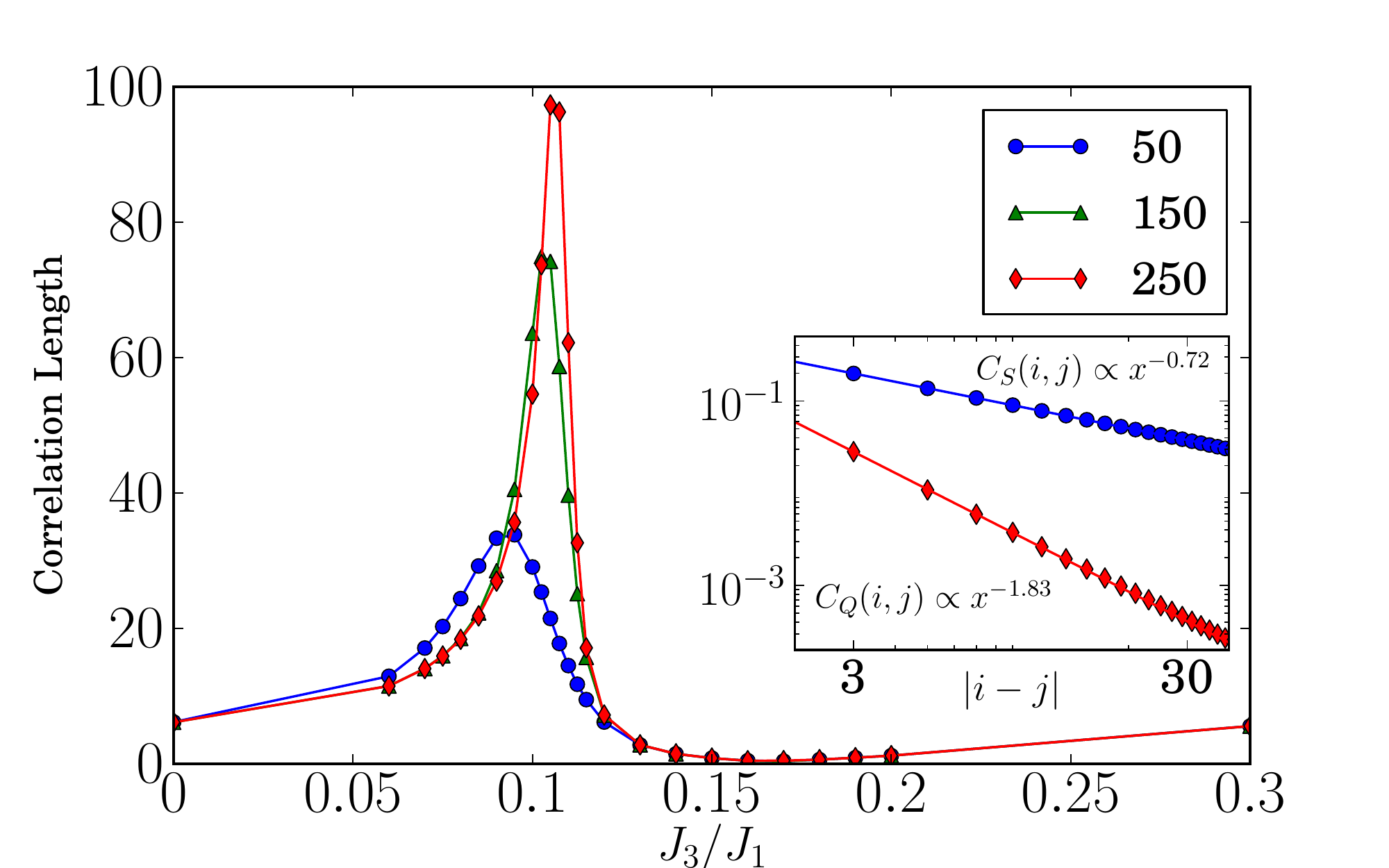}
\caption{(Color online) Correlation length as a function of $J_3/J_1$ for different system sizes. Inset: spin and quadrupolar correlation functions of Eqs.~\eqref{eq:correlationfunction} at the critical point $J_3/J_1 = 0.111$.}
\label{fig:maximum}
\end{figure}

This is further corroborated by our results for the correlation length, which we have obtained by fitting the exponential decay of the spin-spin correlation function with $x^{-1/2}\exp{(-x/\xi)}$.
The results up to 250 sites shown in Fig.~\ref{fig:maximum} are consistent with a divergence at $J_3/J_1 \approx 0.11$. At the MG point, the correlation vanishes rigorously for all sizes.
Together with the results for the dimerization, we therefore conclude that the transition is located at $J_3/J_1 \approx 0.111$. In the Supplemental Material, we also report on a scaling analysis of the fidelity susceptibility \cite{PhysRevE.74.031123} that agrees with this estimate.

\begin{figure}[b]
\includegraphics[width=0.5\textwidth]{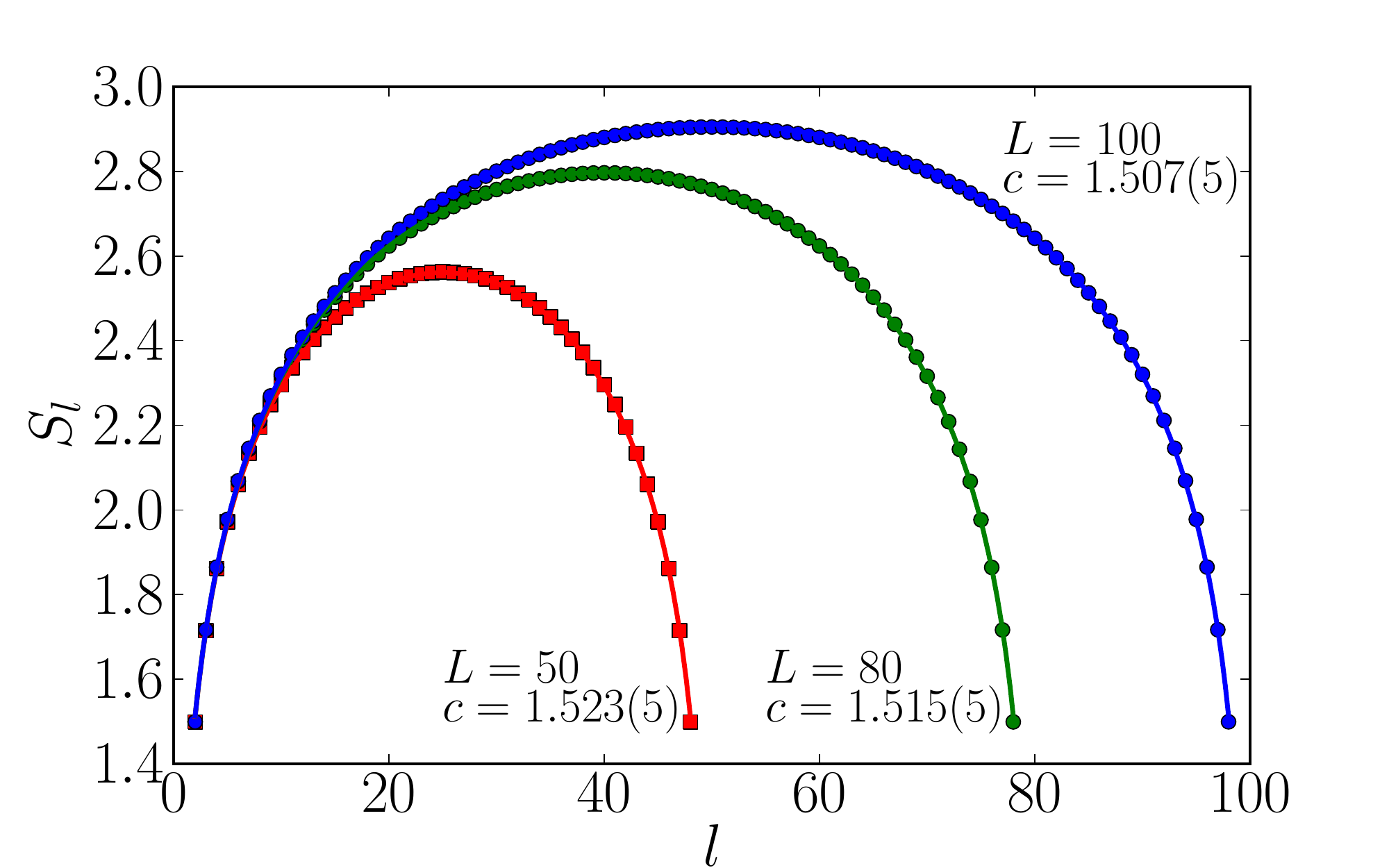}
\caption{(Color online) Fit of the Calabrese and Cardy formula [Eq.~\eqref{eq:entropy}, continuous line] to the DMRG results (dots) for the block entropy at $J_3/J_1 = 0.111$.}
\label{fig:cc}
\end{figure}

Let us now try to further characterize the universality class of this phase transition.
To this end, we have computed the central charge $c$ from the block entropy of the system, $S_\ell = - {\rm Tr} \varrho_\ell \ln \varrho_\ell$, with $\varrho_\ell$ the reduced density matrix of a subsystem of size $\ell$.
For a gapless system,
\begin{equation}
S_\ell = \frac{c}{3} \ln \left[ \frac{L}{\pi} \sin \left( \frac{\pi \ell}{L}\right) \right] + g_{\rm PBC},
\label{eq:entropy}
\end{equation}
in the presence of periodic boundary conditions, so that the central charge $c$ is obtained by fitting the numerical results to Eq.~\eqref{eq:entropy} \cite{Cardy}.
Results for 50, 80 and 100 sites\footnote{The central charge was computed with PBC. We keep up to 2500 states and performed 24 sweeps. This keeps the discarded weight in the last sweep  below  $10^{-7}$} are shown in Fig.~\ref{fig:cc}.
They point rather convincingly to $c=3/2$. This suggests that the transition might be in the SU$(2)_{k=2}$ WZWN universality class \cite{PhysRevB.36.5291}, as the Takhtajan-Babujian (TB) point of the $S=1$ BLBQ chain, at which a
transition from a gapped Haldane phase to a gapped dimerized phase takes place\cite{takhtajan,babujian}.

To further test this conclusion, we have attempted to determine the scaling dimensions at the critical point which determine the exponents of the algebraic decay of the spin and quadrupolar correlation functions \cite{tsvelik_book,PhysRevB.83.184433}
\begin{eqnarray}
C_S(i,j)\equiv\langle S_i^z S_j^z \rangle \sim
(-1)^{i-j}\ (i-j)^{-\frac{1}{4}-\pi/(2\alpha^2)}\ , \label{eq:correlationfunction} \\
C_Q(i,j)\equiv\langle \frac{1}{2}\left(S_i^+\right)^2 \left(S_j^-\right)^2 + h. c. \rangle \sim (i-j)^{-2\pi/\alpha^2}. \nonumber
\end{eqnarray}
For the SU$(2)_{k=2}$ WZWN transition, $\alpha=\sqrt{\pi}$, i.e., the correlation functions decay with exponents 3/4 and 2, respectively.
A fit to the DMRG data at $J_3/J_1 =0.111$ leads to exponents 0.72 and 1.83 for the corresponding correlation functions (see inset of Fig.~\ref{fig:maximum}), in reasonable agreement with the field theory prediction.
Furthermore, the finite-size scaling of the correlation length at the critical point is linear to a very good accuracy, which indicates that $\nu=1$.
Finally, at the critical point we find $d \propto  L^{-0.47}$, implying $\beta/\nu \simeq 0.47$, hence $\beta \simeq 0.47$ since $\nu=1$.
In a related model, Nersesyan and Tsvelik \cite{Nersesyan_Tsvelik_PRL1997} have predicted that the dimerization order parameter can be described as the product of four Ising fields.
Three of them are ordered in the dimerized phase, one is disordered, and they are all critical at the transition point.
Since the Ising exponent $\beta$ is equal to 1/8, we expect the product of four critical Ising fields to scale with exponent $\beta=1/2$.
Again, the numerical estimate is in reasonable agreement with this prediction
\footnote{Note that the same procedure applied at the TB point of the $S=1$ BLBQ chain yields a similar value for the exponent (0.46).}.

\begin{figure}[t]
\includegraphics[width=0.45\textwidth]{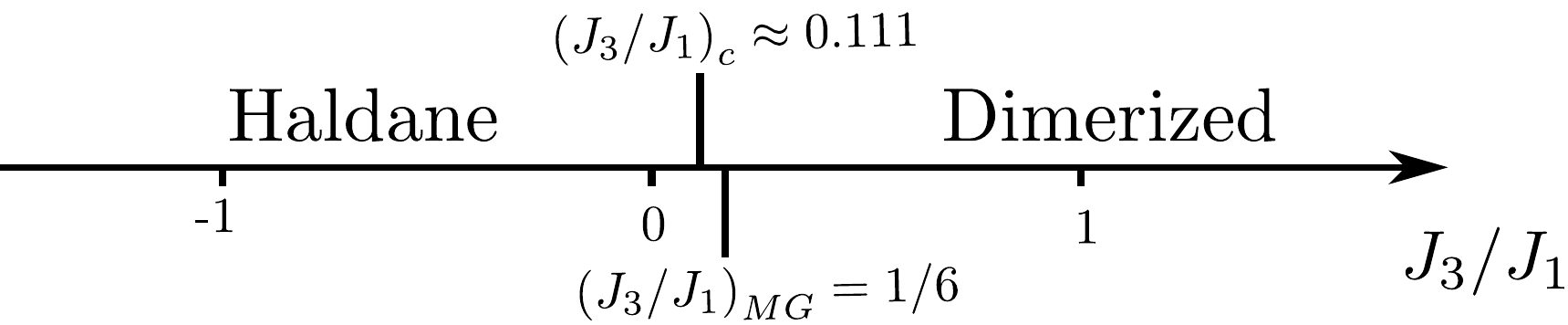}
\caption{Phase diagram of the $J_1 -J_3$ chain of Eq.~\eqref{eq:J1J3ham} for $S=1$.
}
\label{fig:pd}
\end{figure}

We therefore safely conclude that the MG point is representative of an extended phase which is separated from the Haldane phase by a continuous phase transition at $J_3/J_1 \simeq 0.111$, and which extends to large values of $J_3$, as in the $S=1/2$ case \cite{PhysRevB.54.9862}.
The results are summarized in the phase diagram of Fig.~\ref{fig:pd}.

\textit{Discussion -- }
Finally, let us discuss the implications of the present results for actual spin chains.
For simplicity, we concentrate on spin-$1$ chains.\footnote{Similar arguments should
apply to arbitrary $S$ since the fourth-order terms of the strong coupling expansion
have to be the same by symmetry and counting arguments.}
Starting from a two-orbital Hubbard model with repulsion $U$ and Hund's rule coupling,
a strong coupling expansion leads, to second order in the hopping integrals, to the $S=1$ Heisenberg
model with bilinear coupling $J_1$ . At fourth order, three extra terms appear: the three-body interaction $J_3$
of Eq.~\eqref{eq:J1J3ham}, a next-nearest neighbor bilinear coupling $J_2$, as in Eq.~\eqref{eq:mgham},
and a biquadratic interaction  $J_{biq} (\mathbf{S}_i \cdot \mathbf{S}_{i+1})^2$
(see Supplementary Material). The nature of the phase induced by these terms will depend on the
microscopic parameters, but a reasonable case in favor of a spontaneous dimerization in a realistic
parameter range can be articulated around four points: 1) the $J_3$ coupling generated by the fourth
order perturbation theory is essentially always positive; 2) the critical ratio for dimerization
$J_3/J_1=0.111$ is quite small and can be reached for reasonable values of $U$;
3) the biquadratic interaction may be positive or negative. If it is negative, it favors dimerization.
If it is positive, it is typically of the same order as $J_3$, and preliminary results show that it should be significantly larger than $J_3$ to suppress dimerization;
4) to fourth-order, the next-nearest neighbor interaction is essentially ferromagnetic, and this would
compete with dimerization. However, in actual antiferromagnets, it is in fact more likely to be
antiferromagnetic due to residual direct superexchange, hence to be compatible with dimerization.
So, we believe that the dimerization mechanism described by the model of Eq.~\eqref{eq:J1J3ham}
is a realistic potential source of dimerization in actual antiferromagnetic spin chains.
We also note that for systems of ultracold alkaline earth atoms on optical lattices, higher order
perturbation theory leads to the three-body term of Eq.~\eqref{eq:J1J3ham} as well \cite{Gorshkov:2010p1052,gorshkov_communication}.
In actual systems, this dimerization should be observable provided the interchain coupling and the
temperature are both smaller than an energy scale of the order of the gap, a reasonable condition 
since the gap at the Majumdar-Ghosh point is expected to be a significant fraction of $J_1$ (see Supplementary
Material for a detailed discussion). 

\textit{Conclusions --}
We have shown that it is possible to generalize the spin-1/2 $J_1-J_2$ model to larger spins
in such a way that a Majumdar-Ghosh point where dimerized states are exact ground states is
still present without making the model unrealistically complicated. For spin 1, the additional
interaction is a three-site term that appears naturally at fourth order in a 1/U expansion
of a two-band Hubbard model, and we have also shown that the MG point is representative
of an extended dimerized phase separated from the Haldane phase by a continuous transition
in the SU$(2)_{k=2}$ WZWN universality class. We hope that this new model will motivate the
search for experimental realizations in quantum magnets and cold atoms.

\textit{Acknowledgements --} We acknowledge useful discussions with F. Essler, A. V. Gorshkov, K. Penc, A. M. Rey, T. Toth, and S. Wenzel. SRM acknowledges funding by PIF-NSF (grant No. 0904017). This work was supported
by the SNF and by MaNEP.

\newpage

\textbf{Supplemental material to ``Antiferromagnetic spin-$S$ chains with exactly dimerized ground states''}

\section{Fidelity susceptibility for the $S=1$ $J_1-J_3$ chain in the vicinity of the phase transition}

\begin{figure}[b]
\includegraphics[width=0.45\textwidth]{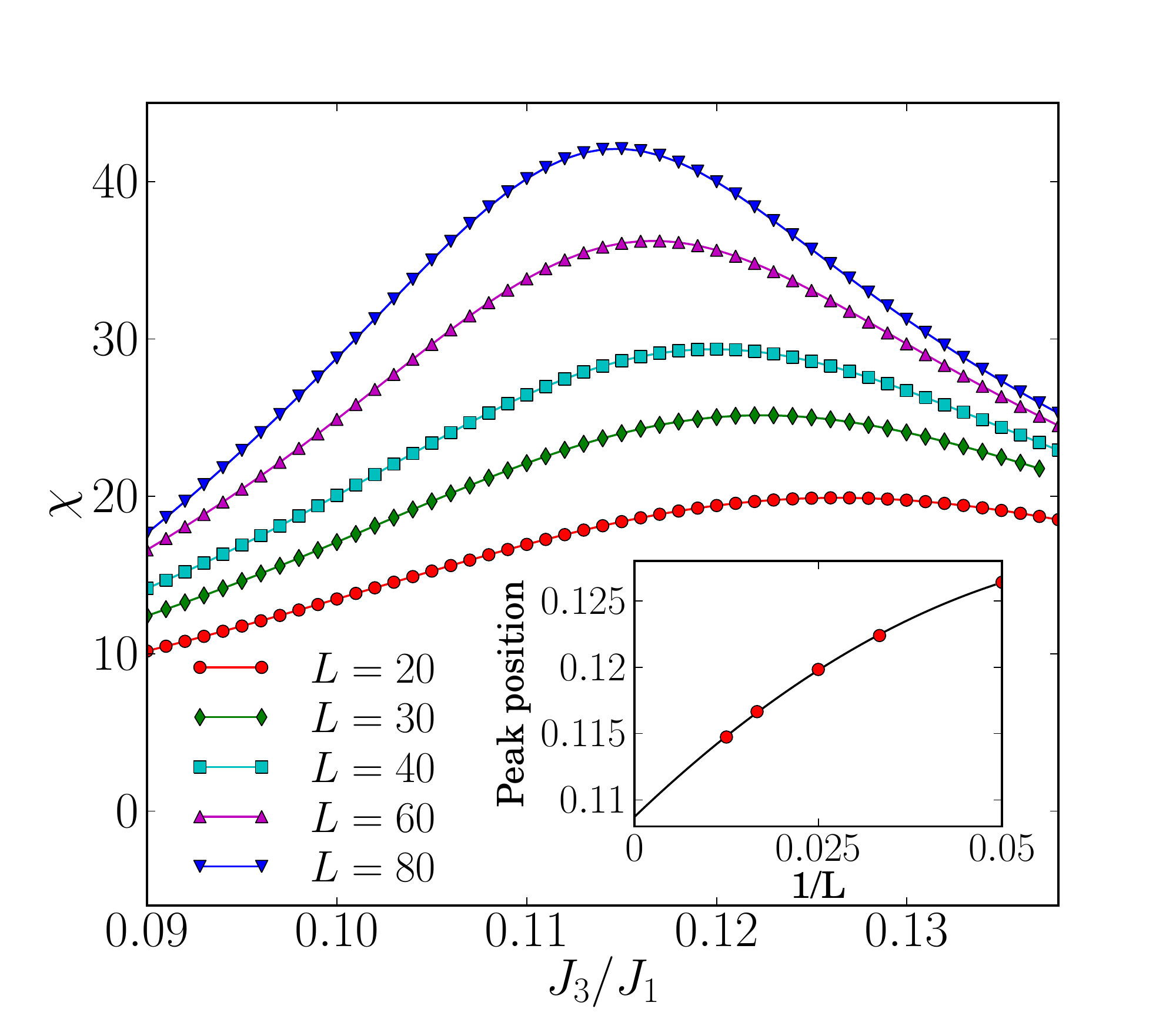}
\caption{(Color online) Fidelity susceptibility [Eq.~\eqref{eq:fsusc}] as a function of $J_3 / J_1$ for different system sizes in the vicinity of the phase transition $J_3 / J_1 \approx 0.11$. The inset shows the finite-size scaling of the peak position.}
\label{fig:fsusc}
\end{figure}

In this section, we discuss in more detail our findings for the fidelity susceptibility \cite{PhysRevE.74.031123} ($J_1 \equiv 1$ in the following)
\begin{equation}
\chi(J_3) = 2 \,  \frac{1 - \left| \langle \psi_0(J_3) | \psi_0(J_3+\delta J_3) \rangle \right|}{L \, (\delta J_3)^2},
\label{eq:fsusc}
\end{equation}
with $|\psi_0(J_3) \rangle$ the ground state of the $J_1-J_3$ chain.
According to the analysis of Refs.~\onlinecite{PhysRevLett.99.095701,PhysRevB.76.180403} and the numerical findings of Refs.~\onlinecite{PhysRevA.77.012311,PhysRevB.77.245109,PhysRevB.78.115410,PhysRevE.76.022101,PhysRevA.84.043601}, in the thermodynamic limit $\chi(J_3)$ should either possess a divergence or a minimum at the critical point, depending on the values of the scaling dimensions and of the critical exponents.
The results of Fig.~\ref{fig:fsusc} indicate that in the present case a peak develops.
We perform an extrapolation of the peak with system size and find that in the thermodynamic limit indeed a divergence is obtained.
As can be seen in the inset of Fig.~\ref{fig:fsusc}, extrapolating the peak position leads to a value $J_c \approx 0.11$, in agreement with the findings for the dimerization and for the correlation length.

\section{Finite size results for the central charge}
To characterize the transition, we have computed the central charge $c$ from the block entropy of the system, $S_\ell = - {\rm Tr} \varrho_\ell \ln \varrho_\ell$, with $\varrho_\ell$ the reduced density matrix of a subsystem of size $\ell$.
For a gapless 1D system with periodic boundary conditions, it behaves as \cite{Cardy}
\begin{equation}
S_\ell = \frac{c}{3} \ln \left[ \frac{L}{\pi} \sin \left( \frac{\pi \ell}{L}\right) \right] + g_{\rm PBC}.
\label{eq:entropy}
\end{equation}

\begin{figure}
\includegraphics[width=0.45\textwidth]{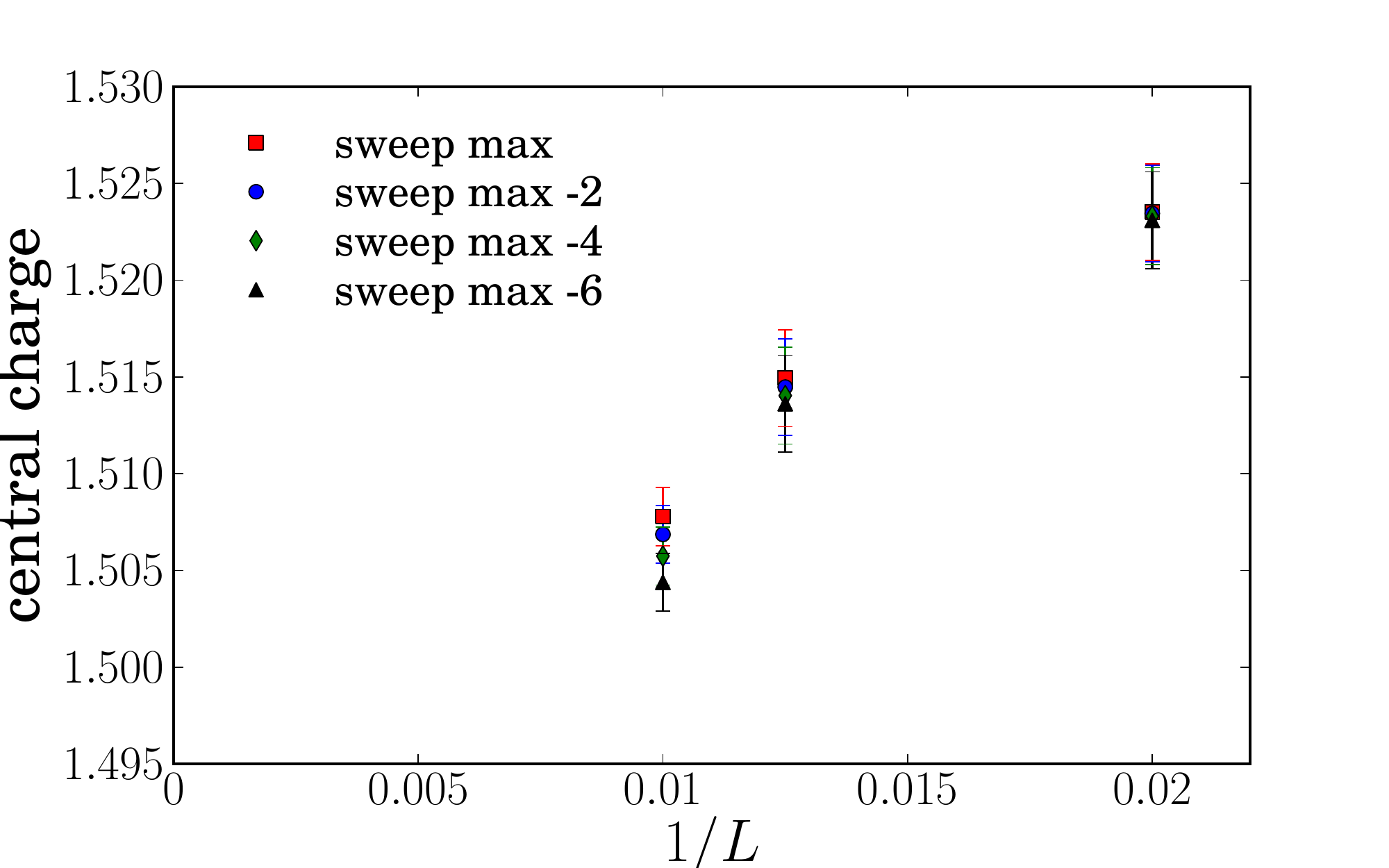}
\caption{(Color online)
Central charge obtained for systems of size $L=50$, $80$ and $100$ from a fit to Eq. (\ref{eq:entropy}) after different numbers of sweeps of the DMRG algorithm. The error bars only take into account the precision of the fit for a given
number of sweeps. As one can see from the evolution of the value of the central charge with the number of sweeps, the values for $L=80$ and $L=100$ have not fully converged due the truncation of the Hilbert space, and the values
for the largest number of sweeps we could achieve are probably underestimates. Because of these uncertainties,
a meaningful finite-size scaling is not possible. Still, these results are clearly compatible with a central charge $c=3/2$.}
\label{fig:c_scaling}
\end{figure}

Due to the limitations of the DMRG when dealing with systems with PBC, we are restricted to system sizes of the order of 100 lattice sites.
In Fig.~\ref{fig:c_scaling}, we show the results for systems with $L=50$, $80$ and $100$ lattice sites including
the error bars that come from the fit to the Calabrese-Cardy formula of Eq. (\ref{eq:entropy}). Another source of error comes from the truncation of the Hilbert space
in the DMRG algorithm. To give the reader an idea of this error, we have plotted the central charge obtained after
different numbers of DMRG sweeps up to the maximal number of sweeps we could achieve. On the very fine scale of the plot, the change in the central charge is negligible for 50 sites, but it is already noticeable for 80 sites and quite significant for 100 sites. Given these uncertainties, and the smallness of the deviations from c=3/2, it does not appear meaningful to perform a finite size extrapolation. This sould be contrasted to the spin-1/2 case treated in \cite{Nishimoto}, where a finite-size analysis could be performed thanks to the good convergence achieved up to 144 sites.
Moreover, one should keep in mind that the critical ratio $J_3/J_1$ is not known exactly, another potential source of error. Still, the results clearly point to a central charge $c=3/2$.

\section{Gap and condition on interchain coupling and on temperature}

\begin{figure}
\includegraphics[width=0.45\textwidth]{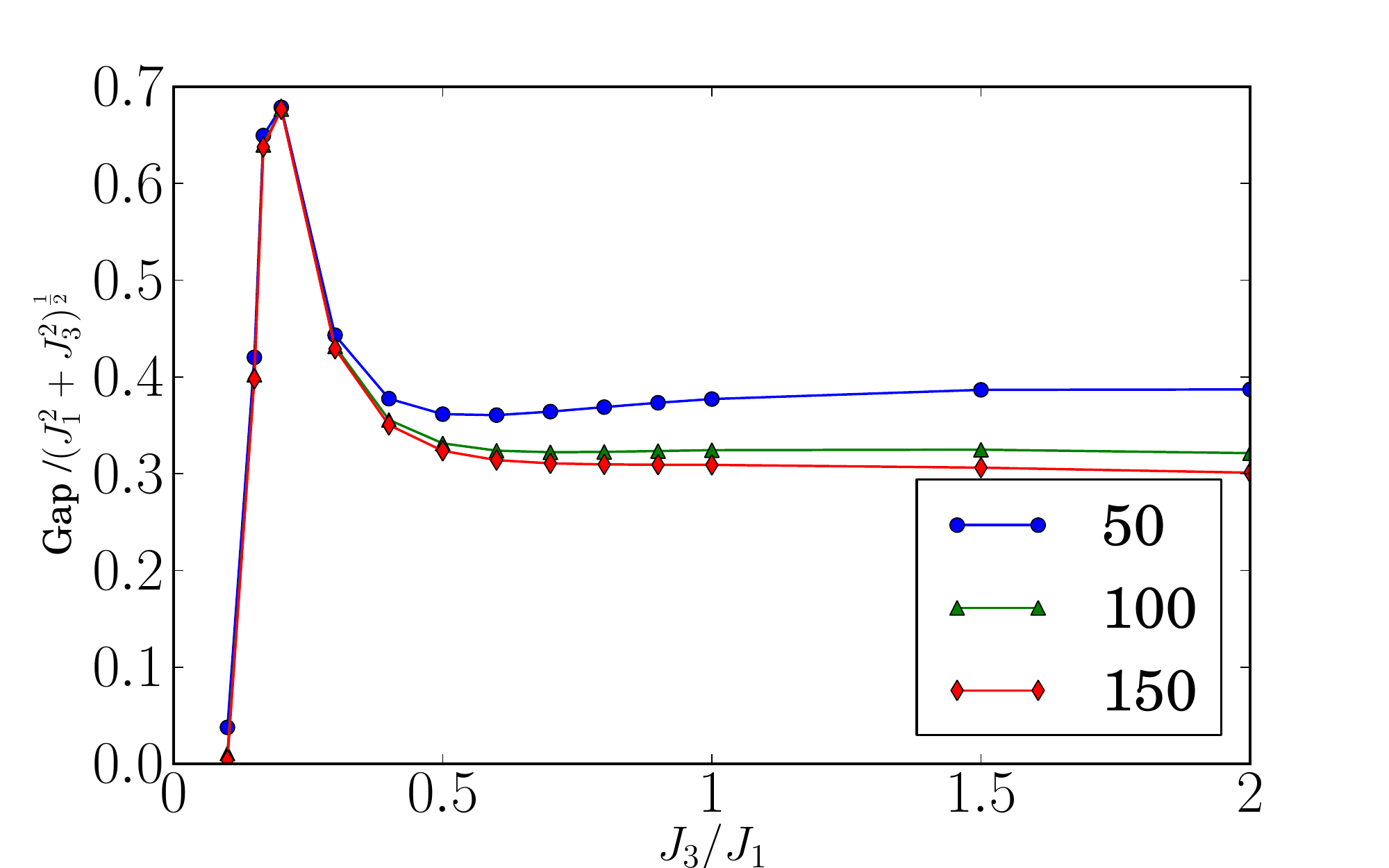}
\caption{(Color online) Gap of the system as a function of $J_3 / J_1$ for different system sizes. }
\label{fig:gap}
\end{figure}

At the Majumdar-Ghosh point, as in the case of the $J_1-J_2$ model, the gap is expected to
be a significant fraction of $J_1$ \cite{white_affleck}. We have checked this expectation with DMRG in the $S=1$ case.
Indeed, as shown in Fig. \ref{fig:gap}, the gap increases very fast above $J_3/J_1=0.11$ to reach values
of the order of $0.7J_1$ around the MG point, above which it decreases to stabilize around
$0.3\sqrt{J_1^2+J_3^2}$.

In antiferromagnets, the conditions to observe the dimerization are thus expected to be met
in sufficiently anisotropic systems and provided one can reach temperatures smaller than the
dominant coupling constant, which is essentially always the case.

In cold atoms, to suppress the interchain coupling, one should work in a 1D trap.
The condition on the temperature should be translated into a condition on entropy. It is not possible
to be quantitative without calculating the temperature dependence of the entropy, a calculation
far beyond the scope of the present analysis, but to reach a temperature equal to a fraction of
the main coupling constant means to work with an entropy per site equal to a fraction of $\ln (2S+1)$.
This is the typical condition to observe antiferromagnetic correlations in cold-atom realizations of the
Heisenberg model, and this is an issue on which the experimental community is currently actively working.

\section{Derivation of the $J_1-J_3$ model from a two-orbital Hubbard model}
The present section is devoted to the derivation of the effective $J_1-J_3$ model from
a microscopic Hubbard model.
In solid state physics spin-1 systems may be realized in the case of transition metal
compounds: If there are 2 degenerate orbitals due to the crystal field splitting and if the
Hund's coupling $J_h$ and the Coulomb repulsion are large enough,
at half-filling (i.e. two spin-1/2 per transition metal ion) the system consists of
localized spin S=1 moments.

Assuming this to be the case, we derive an effective spin S=1
Hamiltonian up to fourth order in degenerate perturbation theory
in the strong coupling limit of a two-orbital Hubbard model on a chain,
and we show under which conditions on the original microscopic parameters (hopping integrals,
Coulomb repulsion, Hund's coupling) this effective Hamiltonian reduces
to a $J_1-J_3$ model with nearest-neighbor coupling and three-body interaction terms.

\subsection{Generalized Hubbard model}
Our starting point is the following Hamiltonian at half-filling:
\begin{eqnarray}
\mathcal{H}_{Hb} &=&  \sum_{i,j}  \sum_{m,m'}\sum_{\sigma} t_{m,m'}^{ij}c_{i m \sigma}^{\dagger}c_{j m' \sigma}  \\
&+&\frac{1}{2}\sum_{i}\sum_{m,m'}\sum_{\sigma ,\sigma'}U_{m m'}n_{i  m  \sigma}n_{i  m' \sigma'} \nonumber \\
&+&\frac{1}{2}\sum_{i}\sum_{m\neq m'}\sum_{\sigma\neq\sigma'}\{J_{h}n_{im\sigma}n_{im'\sigma} \nonumber \\
&+&J_{h}c_{im\sigma}^{\dagger}c_{im\sigma'}c_{im'\sigma'}^{\dagger}c_{im'\sigma} \nonumber \\
 &+& 2 J_{p}c_{im'\sigma'}^{\dagger}c_{im'\sigma}^{\dagger}c_{im\sigma'}c_{im\sigma}\}\nonumber
\label{eq:2bandHM}
\end{eqnarray}

where $i,j$ are the site indices, $m,m'$ refer to the orbitals $a$ and $b$,
and $\sigma$ to the electronic spin. The hopping integrals between
two neighboring orbitals are denoted by $t_{m,m'}^{ij}$, the on-site Coulomb repulsions
by $U_{mm'}$, $J_{h}$ represents the Hund's coupling
and $J_{p}$ the pair hopping amplitude. Furthermore,
we assume that additional relations, typical of cubic symmetry, are
satisfied, namely $U_{aa}=U_{bb}$ and $U=U_{aa}-2J_{h}$. This kind of
Hamiltonian has been extensively discussed in the context of systems with
orbital degeneracy \cite{fazekas,castellani}.

\subsection{Effective spin-model to fourth order in degenerate perturbation theory}
Using degenerate perturbation theory, the effective spin model of Eq.~\eqref{eq:2bandHM}
on a chain takes the form
\begin{eqnarray}
H&=&H^{(2)}+H^{(4)} \nonumber \\
H^{(2)} &=& J^{(2)}_{\text{Heis}} \sum_{<i,j>} ({\bf S}_i \cdot {\bf S}_j) \nonumber \\
H^{(4)}&=&  \sum_{<i,j>}\left( J^{(4)}_{\text{Heis}}{\bf S}_i \cdot {\bf S}_j + J^{(4)}_{\text{Biqu}}({\bf S}_i \cdot {\bf S}_j)^2\right) \nonumber \\
&+&  \sum_{\ll i,j \gg}  J^{(4)}_{\text{nn}} ({\bf S}_i\cdot{\bf S}_j)  \nonumber \\
&+& \sum_{<i,j,k>}J^{(4)}_{\text{3}}  \left[({\bf S}_i\cdot {\bf S}_j)({\bf S}_j\cdot{\bf S}_k)+h.c. \right] , \nonumber \\
\label{eq:spinHam}
\end{eqnarray}
where ${\bf S}$ are spin-1 operators. The sum over $<i,j>$ runs over nearest-neighbor
pairs, the one over $\ll i,j\gg$ runs over next-nearest neighbor pairs, while the one over $<i,j,k>$ runs over all sequences of three spins.

The effective Hamiltonian consists of different terms. At lowest order, one recovers the
Heisenberg model with a nearest-neighbor spin coupling $J^{(2)}_{\text{Heis}}$ which gets
renormalized at fourth order by the coefficients $J^{(4)}_{\text{Heis}}$.
At fourth order, two other 2-site terms appear: A next-nearest neighbor coupling $J^{(4)}_{\text{nn}}$, as
in the S=$\frac{1}{2}$ case \cite{Delannoy}, and a biquadratic coupling $J^{(4)}_{\text{Biqu}}$
typical of $S=1$ systems.
Finally, there is in additional three-site interaction $J^{(4)}_{\text{3}}$ which cannot be reformulated
as a 2-site operator. Let us mention that these terms, which appear in the perturbation expansion
of the Hubbard model, have also been extracted from ab initio calculations
in a different context \cite{Graaf}.

In order to have more compact expressions, we introduce the following relations:
\begin{eqnarray*}
\frac{t_{aa}^4+t_{bb}^4}{2}&=&t_1^4\\
\frac{t_{aa}^2t_{ab}^2+t_{bb}^2t_{ab}^2}{2}&=&t_2^4\\
t_{aa}^2t_{bb}^2&=&t_{2p}^4\\
t_{aa}t_{bb}t_{ab}^2&=&t_4^4\\
\end{eqnarray*}
The various coefficients of \eqref{eq:spinHam} can then be expressed in terms of the microscopic
parameters of the original Hubbard model as sums of terms classified according to the combination
of $U$, $J_h$ and $J_p$ that appears in the denominator:
\begin{widetext}
\begin{eqnarray*}
J^{(2)}_{\text{Heis}} &= &\frac{\text{t}_{aa}^2+2 \text{t}_{ab}^2+\text{t}_{bb}^2}{2 \text{J}_h+U} \\
J^{(4)}_{\text{Heis}} &=&
\frac{-8 \text{t}_1^4-32 \text{t}_2^4-12 \text{t}_{2p}^4+8 \text{t}_4^4-20 \text{t}_{ab}^4}{(2 \text{J}_h+U)^3}+
\frac{2 \text{t}_1^4-2 \text{t}_{2p}^4}{\text{J}_h (2 \text{J}_h+U)^2}+
\frac{-16 \text{t}_2^4-16 \text{t}_4^4}{(-3 \text{J}_h-2 \text{J}_p) (2 \text{J}_h+U)^2}\\
\\&+&
\frac{-8 \text{t}_{2p}^4+16 \text{t}_4^4-8 \text{t}_{ab}^4}{(-4 \text{J}_h-3 U) (2 \text{J}_h+U)^2}+
\frac{-4 \text{t}_1^4+8 \text{t}_4^4-4 \text{t}_{ab}^4}{(-4 \text{J}_h-U) (2 \text{J}_h+U)^2}+\\&+&
\frac{-16 \text{t}_2^4-4 \text{t}_{2p}^4-8 \text{t}_4^4-4 \text{t}_{ab}^4}{(-5 \text{J}_h-2 \text{J}_p-U) (2 \text{J}_h+U)^2}+
\frac{-4 \text{t}_{2p}^4+8 \text{t}_4^4-4 \text{t}_{ab}^4}{(-5 \text{J}_h+2 \text{J}_p-U) (2 \text{J}_h+U)^2}
\\
J^{(4)}_{\text{Biqu}}&=&
\frac{2 \text{t}_1^4+8 \text{t}_2^4+2 \text{t}_{2p}^4+4 \text{t}_{ab}^4}{(2 \text{J}_h+U)^3}+
\frac{-16 \text{t}_2^4-16 \text{t}_4^4}{(-3 \text{J}_h-2 \text{J}_p) (2 \text{J}_h+U)^2}+
\frac{16 \text{t}_2^4-16 \text{t}_4^4}{(-5 \text{J}_h-2 \text{J}_p) (2 \text{J}_h+U)^2}+\\&+&
\frac{\frac{3 \text{t}_1^4}{2}+2 \text{t}_2^4-
\frac{5 \text{t}_{2p}^4}{2}-\text{t}_{ab}^4}{\text{J}_h (2 \text{J}_h+U)^2}+
\frac{4 \text{t}_{2p}^4-8 \text{t}_4^4+4 \text{t}_{ab}^4}{(-6 \text{J}_h+4 \text{J}_p) (2 \text{J}_h+U)^2}+\\&+&
\frac{4 \text{t}_{2p}^4+8 \text{t}_4^4+4 \text{t}_{ab}^4}{(-6 \text{J}_h-4 \text{J}_p) (2 \text{J}_h+U)^2}+
\frac{-2 \text{t}_{2p}^4+4 \text{t}_4^4-2 \text{t}_{ab}^4}{U (2 \text{J}_h+U)^2}
\\
J^{(4)}_{\text{3}} &=&
\frac{4 \text{t}_1^4+16 \text{t}_2^4+2 \text{t}_{2p}^4+4 \text{t}_4^4+6 \text{t}_{ab}^4}{(2 \text{J}_h+U)^3}+
\frac{\text{t}_1^4-\text{t}_{2p}^4}{\text{J}_h (2 \text{J}_h+U)^2}+
\frac{-8 \text{t}_2^4-8 \text{t}_4^4}{(-3 \text{J}_h-2 \text{J}_p) (2 \text{J}_h+U)^2}+\\&+&
\frac{4 \text{t}_{2p}^4-8 \text{t}_4^4+4 \text{t}_{ab}^4}{(-2 \text{J}_h-3 U) (2 \text{J}_h+U)^2}+
\frac{2 \text{t}_1^4-4 \text{t}_4^4+2 \text{t}_{ab}^4}{(-4 \text{J}_h-U) (2 \text{J}_h+U)^2}+
\frac{8 \text{t}_2^4+2 \text{t}_{2p}^4+4 \text{t}_4^4+2 \text{t}_{ab}^4}{(-5 \text{J}_h-2 \text{J}_p-U) (2 \text{J}_h+U)^2}+\\&+&
\frac{2 \text{t}_{2p}^4-4 \text{t}_4^4+2 \text{t}_{ab}^4}{(-5 \text{J}_h+2 \text{J}_p-U) (2 \text{J}_h+U)^2}
\end{eqnarray*}
\begin{eqnarray*}
J^{(4)}_{\text{nn}} &=&\frac{-4 \text{t}_1^4-16 \text{t}_2^4-8 \text{t}_4^4-4 \text{t}_{ab}^4}{(2 \text{J}_h+U)^3}+
\frac{-2 \text{t}_1^4+2 \text{t}_{2p}^4}{\text{J}_h (2 \text{J}_h+U)^2}+
\frac{16 \text{t}_2^4+16 \text{t}_4^4}{(-3 \text{J}_h-2 \text{J}_p) (2 \text{J}_h+U)^2}+\\&+&
\frac{-4 \text{t}_{2p}^4+8 \text{t}_4^4-4 \text{t}_{ab}^4}{(-2 \text{J}_h-3 U) (2 \text{J}_h+U)^2}+
\frac{-2 \text{t}_1^4+4 \text{t}_4^4-2 \text{t}_{ab}^4}{(-4 \text{J}_h-U) (2 \text{J}_h+U)^2}+
\frac{-8 \text{t}_2^4-2 \text{t}_{2p}^4-4 \text{t}_4^4-2 \text{t}_{ab}^4}{(-5 \text{J}_h-2 \text{J}_p-U) (2 \text{J}_h+U)^2}+\\&+&
\frac{-2 \text{t}_{2p}^4+4 \text{t}_4^4-2 \text{t}_{ab}^4}{(-5 \text{J}_h+2 \text{J}_p-U) (2 \text{J}_h+U)^2}.
\end{eqnarray*}
\end{widetext}

\end{document}